# Graphene plasmonics – optics in flatland

A. N. Grigorenko[1*], M. Polini[2], K. S. Novoselov[1]

[1]School of Physics and Astronomy, University of Manchester, Manchester, M13 9PL, UK

[2]NEST-CNR-INFM and Scuola Normale Superiore, I-56126 Pisa, Italy

**Abstract:** Two rich and vibrant fields of investigation – graphene and plasmonics – strongly overlap. Not only does graphene possess intrinsic plasmons that are tuneable, adjustable and have relatively low dissipation, but also a combination of graphene with noble-metal nanostructures promises a variety of exciting applications for conventional plasmonics. Graphene versatility means that graphene based plasmonics can enable the manufacture of novel optical devices working in different frequency ranges - from terahertz to visible light - with unrivalled speed, low driving voltage, low power consumption and a small physical signature. Here we review the field emerging at the intersection of graphene physics and plasmonics.

*e-mail: sasha@manchester.ac.uk

Graphene – a two-dimensional (2D) layer of carbon atoms arranged in a honeycomb lattice[1,2] – has already demonstrated unique mechanical, electric, magnetic and thermal properties which guarantee a multitude of exciting applications vigorously pursued by academia and industry[3]. Interestingly, it is in optics where graphene has shown its true "colours" and where the first commercial application of graphene has been realised[4]. Graphene has an extremely high quantum efficiency for light-matter interactions and strong optical non-linearity, contains unusual and refreshingly different plasmons, can be modified by gating, doping, chemistry and conventional plasmonics based on noble metals. The optics and photonics of graphene have often been reviewed[4-8]. However, the recent explosion of works on graphene plasmonics and optics of 2D single-layer materials calls for a fresh look at graphene and its remarkable properties.

## Optical properties of graphene and other 2D materials

### Infrared limit

Among many of the unique properties of graphene[3], probably the most peculiar is the fact that quasiparticles in this material obey a linear dispersion relation. As a consequence, an additional– chiral – symmetry exists for the quasiparticles, which fixes the direction of pseudospin (in a given valley) to be parallel or antiparallel to the direction of motion for electrons and holes respectively[9]. The linear spectrum and chirality have an immediate and dominant impact on the electronic and optical properties of this 2D crystal: such quasiparticles cannot be localized (Klein paradox)[10], interference corrections to conductivity are positive (weak antilocalization) and the optical conductivity of graphene is constant and independent of energy. Furthermore, it is easy to demonstrate that the optical conductivity is also independent of any material parameters (apart from the number of species – 4 for graphene) – $\sigma_{uni} = \pi e^2/(2h)$ (here $e$ is electron charge and $h$ is the Plank constant)[11,12], so that the optical absorption depends only on the fine structure constant: $\pi\alpha \approx 2.3\%$ (Fig. 1a). Interestingly, the optical conductivity is close, but not equal to the DC conductivity in the ballistic regime – $4e^2/(\pi h)$[13,14].

Such simple behaviour is expected for undoped samples at zero temperatures. At the same time, the ability to dope graphene either chemically or electrostatically is at the heart of many graphene-based devices[1], and such doping has a very strong effect on the optical properties[15]. Pauli blocking (explained in Fig. 1b) ensures that photons with energy smaller than $2E_F$ ($E_F$ is the Fermi energy for electrons or holes) are not absorbed (here we ignore the asymmetry between valence and conduction bands). This "unaccounted" spectral weight can be linked via the sum rule to the pronounced Drude peak at zero frequency. However, above the Pauli blocking energies the absorption is still given by a simple relation, $\pi\alpha$.

### Visible range

The situation become less simple at energies where the dispersion deviates from simple linear behavior – it becomes anisotropic (trigonal warping) and sub-linear. However, even within the visible range the optical absorption is still close to $\pi\alpha$, so the optical absorption even at 3eV is less than 3%, see Fig. 1a. This property of graphene, in conjunction with very low sheet resistance (can be below 100Ω/sq in doped samples) is a reason that graphene is considered to be a very promising material for a number of applications where a transparent and conductive layer is required (touch screens, liquid crystal displays, organic light emitting diodes displays, solar cells, etc). Recently, gating with a solid electrolyte allowed carrier concentrations as large as $10^{14}$ cm$^{-2}$ to be achieved, which converts into $E_F$~1eV, so a modulation of optical transmission in the visible spectrum is possible[15-17]. Combining graphene with silicon waveguides, for example, allows one to produce a broadband graphene-based waveguide-integrated optical modulator[17] shown in Fig. 1c.

### Ultraviolet limit and excitonic effects

Above 3 eV deviations from the linear dispersion and trigonal warping effects become extremely strong. The band structure of graphene is such that it has saddle points at M points of the Brillouin zone, which lead to van Hove-like singularities. In the single-electron approximation this should result in a strong absorption peak (above 10% - a very appreciable number for a monolayer of atoms) at around 5.2 eV. In practice, this simple picture fails and the peak is observed close to 4.5 eV[18], see Fig. 1a. The strong changes in the absorbance spectrum are mainly due to resonant excitons.

### Graphene in magnetic field

The linear spectrum and large Fermi velocity of graphene lead to the observation of the quantum Hall effect at room temperature[19]. The optical effect that is the analogue of the Hall effect in DC measurements is the Faraday rotation of polarization. It has been demonstrated[20] that graphene, even though only one atom thick, has a very large Faraday rotation of about 6 degrees at modest magnetic fields of a few Tesla. This offers the possibility to use graphene in magneto-optical devices.

### Strained and curved graphene

So far we have mainly concentrated on graphene as a 2D electronic system. However, this material is a true atomic membrane which can be stretched, compressed, folded, etc. As the curvature can be controlled externally, graphene could be used as adaptive lens simply due to geometrical effects[21]. It has also been demonstrated both theoretically and experimentally that the electronic properties of graphene are strongly dependent on local strain[22,23]. Recently it has been predicted that the optical properties of graphene could be modified by applied strain. For instance, in the case of a uniaxial strain[24,25], a strong dichroism has been found, so that the absorption coefficient depends strongly on the applied strain and varies with the relative orientation of the uniaxial strain and the polarization of the incoming light. At extreme polarization angles, the optical absorption can be expressed as $\pi\alpha(1\pm 4\varepsilon)$, where $\varepsilon$ is the uniaxial strain. The effect is even stronger for the UV light, as the van Hove singularities are shifted by more than 1 eV for 10% strain (Fig. 1a).

## Optical properties of other 2D materials

Graphene, possessing a number of nontrivial, interesting and, possibly, useful properties has another one, which is often overlooked – it opened a floodgate, allowing other 2D crystals and heterostructures based on such crystals to be discovered and studied. Indeed, over the last few years a large number of 2D materials have been intensively investigated and their optical properties are exciting and often different from the properties of their three-dimensional parent materials. There is a very large variety of 2D crystals now available, ranging from various modifications of graphene (bilayer, trilayer, chemically functionalized graphene) to complex materials as dichalcogenides and complex oxides. Although the optical properties of a few of these crystals have been studied to date, it is clear that this will become a major field of research in the future.

**Bi- and trilayer graphene: material with controllable bandgap and controllable optical absorption**

The electronic structure of bilayer graphene is dramatically different from that of the monolayer, so it can be considered as a separate 2D material in its own right. Indeed, the band structure of A-B (Bernall) stacked graphene is still gapless with valence and conduction bands having parabolic dispersion and touching at zero energy. It also contains additional sub-bands which, in the first approximation, are offset from zero energy by $\gamma_1 \approx 0.4$ eV (the nearest neighbour hopping between layers). In optics this leads to a strong absorption peak at this energy (Fig. 1a).

The gapless spectra in mono- and bilayer graphene are protected by the symmetry between the sub-lattices. However, in bilayer such symmetry also implies the symmetry between the two layers and can be easily lifted by selective chemical doping of one of the layers or by applying transverse electric field (gating), leading to the opening of a significant gap. Indeed, such a gap has been observed in optical experiments on this material, and similar effects are also responsible for strong changes in the optical spectrum around 0.4 eV. It has been proposed that these effects can be used for optical modulation in telecommunications[26]. Figure 2a shows the energy spectrum for bilayer graphene and Fig. 2b demonstrates an effective optical modulator based on the tunable bandgap in bilayer graphene[26].

Trilayer graphene is another interesting material, which comes in two very different configurations. The electronic structure of Bernal stacked (A-B-A) can be viewed as a combination of those from one mono- and one bi-layer graphene sheets. At the same time, rhombohedral (A-B-C stacked) trilayer more resembles bilayer graphene. As a result, the optical properties of both are strongly dependent on the perpendicular electric field, with the A-B-C trilayer demonstrating a larger band gap opening than A-B-A. As the van Hove singularities in this material are stronger than in bilayer graphene, so are the changes in optical absorption, which might be of interest for applications which requiring a tuneable bandgap.

**Graphane and Fluorographene**

Graphene can also be considered as a giant aromatic molecule which can be subjected to chemical reactions. Indeed two such chemical modifications have been obtained: graphane[27] (when a hydrogen atom is attached to each of the carbon atoms) and fluorographene[28] (or 2D *Teflon*, where one fluorine atom is attached to each carbon atom). Covalently bonded hydrogen or fluorine changes the hybridization of carbon atoms from $sp^2$ to $sp^3$, which removes the π orbitals from the electronic band structure. This should lead to the opening of a large gap of the order of the separation between the *σ* bands. Experimentally, optical gaps of the order of 3 eV have been observed[28] for fluorographene (Fig. 1a), which probably indicates the presence of a large number of defects. This shows that optical properties of 2D crystals can be modified via chemical modification of the materials.

**2D atomic crystals and their heterostructures**

Monolayer molybdenum disulphide is probably the second most studied 2D material after graphene. With both synthetic and natural crystals available, $MoS_2$ has been exfoliated to the monolayer state by both mechanical[29] and liquid phase[30] exfoliation. Perhaps surprisingly, the properties of monolayer $MoS_2$ are radically different from the properties of the 3D material. Bulk crystals have an indirect band gap of the order of 1.29 eV. Once in the 2D state, $MoS_2$ exhibits a direct band gap[31] of the order of 1.9 eV around the K points of the Brillouin zone, which leads to a strong increase in luminescence[32], see Fig. 1a.

Furthermore, due to the absence of the inversion symmetry, the strong spin-orbit interaction splits the valence bands at the K and K' points by about 0.16 eV. In addition, time reversal symmetry ensures that this splitting is in opposite directions in the two valleys, leading to the opposite spin polarization in the K and K' valleys. Effectively, this allows optical control (by means of circularly polarized light) of the population of charge carriers separately in each of the valley[33], leading to the realization of *"valleytronics"*.

The electronic and optical properties of other 2D crystals (BN[34], TaS$_2$, NbSe$_2$, WS$_2$, *etc*) are currently being intensively investigated. However, one can anticipate more surprises from heterostructures created by stacking such 2D crystals one on top of another[35]. Even the simplest possible stacks have already revealed new physics[36-38] and could revolutionize some applications. The optical properties of these heterostructures can be tuned very accurately, as materials with very different bandgaps and thicknesses can be combined, see Fig. 2c for illustration. In particular, such heterostructures could be used for photodetection, solar-harvesting, optical signal modulation, etc. Another direction is to combine graphene heterostructures with standard optics (e.g., optical cavities[39]) and optoelectronic devices (e.g., integrated interferometers[5]) as Fig. 2d demonstrates.

## Intrinsic graphene plasmons

Plasmons are ubiquitous high-frequency collective density oscillations of an electron liquid, which occur in many metals and semiconductors[40]. The intrinsic graphene plasmons are refreshingly different from plasmons in noble metals as they can be tuned by gating or doping, do not exhibit large Ohmic losses and can be confined to tighter regions[6]. Graphene plasmonic resonances could play a pivotal role in the realization of robust and cheap photodetectors of Terahertz (THz) radiation, important for homeland security. We consider the peculiar properties of the intrinsic plasmon modes of the electron gas in a pristine graphene sheet. Our focus is on the plasmons of doped samples, although interesting collective modes have also been predicted for undoped graphene[41]. Moreover, we will focus on longitudinal modes, i.e. modes whose associated electric field is parallel to the wave vector *q*. The existence of a transverse collective mode in graphene (with a frequency slightly lower than the Pauli-blocking threshold for interband absorption) has been also discussed[42].

### Electron plasma in 2D

Two-dimensional electron systems (2DESs) have been a fertile source of exciting physics for more than four decades. As in any 2DES, electrons in graphene do not move as independent particles. Rather, their motions are highly correlated due to pairwise interactions. These are described by a potential $u(r_{ij}) = u(|\mathbf{r}_i - \mathbf{r}_j|)$, which depends only on the absolute value of the relative distance $\mathbf{r}_{ij} = \mathbf{r}_i - \mathbf{r}_j$ between two electrons. The interaction potential is sensitive to the dielectric media surrounding the graphene sheet. For graphene with one side exposed to a medium with dielectric constant $\varepsilon_1$ and the other to one with dielectric constant $\varepsilon_2$ we have: $u(r_{ij}) = e^2/(\varepsilon r_{ij})$, where $\varepsilon = (\varepsilon_1 + \varepsilon_2)/2$ (so that the Fourier transform is $u_q = 2\pi e^2/(\varepsilon q)$). The electron gas in a graphene sheet can be described at low energies by the following continuum-model Hamiltonian:

$$\hat{H} = v_F \sum_i \boldsymbol{\sigma} \cdot \mathbf{p}_i + \frac{1}{2} \sum_{i \neq j} \frac{e^2}{\varepsilon |\mathbf{r}_i - \mathbf{r}_j|}, \qquad (1)$$

where $v_F \approx 10^6$ m/s is the Dirac velocity, $\mathbf{p}_i = -i\hbar \nabla_{\mathbf{r}_i}$ is the canonical momentum of the *i*-th electron, and $\boldsymbol{\sigma} = (\sigma_x, \sigma_y)$ are 2D vectors of the Pauli matrices. For the sake of simplicity, eq. (1) has been written for a single-channel massless Dirac fermion (MDF) model, i.e. it holds for electrons with given spin and valley indices. The relative importance of electron-electron interactions is quantified by the ratio between the "magnitude" of the second term with respect to that of the first term. For a doped graphene sheet, the typical distance between electrons is of the order of the inverse of the Fermi wave number, i.e. $k_F^{-1}$. The second term is thus of the order of $e^2 k_F / \varepsilon$. The kinetic energy is of the order of $\hbar v_F k_F$ and hence the ratio between these two quantities defines a dimensionless parameter, $\alpha_{ee}$, usually termed "graphene's fine-structure constant":

$$\alpha_{ee} = \frac{e^2}{\varepsilon \hbar v_F}. \qquad (2)$$

It is easy to see that $\alpha_{ee}$ can be expressed using the fine structure constant $\alpha = e^2/(\hbar c)$ as $\alpha_{ee} = \frac{1}{\varepsilon} \frac{c}{v_F} \alpha$. This allows us to quickly estimate the magnitude of $\alpha_{ee}$. For example, for graphene with one side exposed to air ($\varepsilon_1 = 1$) and the other to SiO$_2$ ($\varepsilon_1 \cong 3.9$), we have $\alpha_{ee} \cong 0.9$. For a suspended graphene sheet $\varepsilon_1 = \varepsilon_2 = 1$ and $\alpha_{ee} \cong 2.2$. Hence, the graphene fine-structure constant can be tuned experimentally by changing the dielectric environment surrounding the graphene[43]. From this analysis we conclude that, at least in principle, electrons in a doped graphene sheet interact quite strongly with each other and that interaction effects in this material have to be analyzed with great care.

**Theory of 2D plasmons**

The physical origin of plasmons can be understood as follows. When electrons move to screen an electric field, they tend to overshoot the mark, see Fig. 3a. They are then pulled back toward the charge disturbance and overshoot again, setting up a weakly damped oscillation. The restoring force responsible for the oscillation is proportional to the gradient of the self-consistent field created by all the electrons. The plasmon dispersion in a 2DES can be understood quite easily, at least in the long-wavelength $q \ll k_F$ limit. In this limit one can use a macroscopic description for the electron dynamics introducing two collective variables[40], which obey macroscopic conservation laws: the deviation of the electron density $\delta n(\mathbf{r},t)$ and the associated current density $\delta \mathbf{j}(\mathbf{r},t)$. For $|\delta n / n| \ll 1$ the plasmons are described by linearized Euler equation of motion

$$\frac{\partial \mathbf{j}(\mathbf{r},t)}{\partial t} = -\frac{D}{\pi e^2} \nabla_{\mathbf{r}} \int d^2\mathbf{r}' \frac{e^2}{\varepsilon |\mathbf{r} - \mathbf{r}'|} \delta n(\mathbf{r}',t), \qquad (3)$$

which, in conjunction with the continuity equation, leads the following equation for the Fourier component of $\delta n(\mathbf{r},t)$

$$\left(\omega^2 - \frac{D}{\pi e^2} q^2 u_q \right) \delta n(\mathbf{q},\omega) = 0, \quad (4)$$

where $D$ is the so-called "Drude weight". Eq. (4) implies the existence of plasmons with a frequency $\omega_{pl}(q) = \sqrt{2Dq/\varepsilon}$. Notice that $\omega_{pl}(q) \propto \sqrt{q}$, a peculiarity of plasmon oscillations in 2D. For ordinary parabolic-band fermions with mass $m_b$ the Drude weight is given by $D = \pi e^2 n / m_b$ and we have the well-known result: $\omega_{pl}(q) = \sqrt{2\pi n e^2 q / (\varepsilon m_b)}$. The situation turns out to be quite different for MDF in graphene. The Drude weight of non-interacting MDFs is given by $D_{MDF} = 4 E_F \sigma_{uni} / \hbar$ which yields the plasmon frequency in doped graphene[44-47] (for $E(q) = \hbar \omega_{pl}(q) < 2 E_F$)

$$\omega_{pl}(q) = \sqrt{\frac{8 E_F \sigma_{uni}}{\hbar \varepsilon} q}. \quad (5)$$

The Dirac plasmon frequency thus scales like $E_F^{1/2} \propto n^{1/4}$ and contains the Planck's constant. For $E(q) > 2 E_F$ interband transitions make graphene plasmons strongly dissipative. For a typical doping $n$ = $10^{11}$ cm$^2$, the Fermi energy in graphene is $E_F$~37 meV and the plasmon energy for graphene on SiO$_2$ at $q = 0.1 k_F$ ~ $0.6 \times 10^5$ cm$^{-1}$ is about 16 meV, which is in the infrared range.

One remarkable conclusion of eq. (5) is the fact that the compression of the surface plasmon wavelength over the excitation wavelength is governed by the fine structure constant and can be strong[6] since $\lambda_{pl} / \lambda_0 \approx 2\alpha E_F / (\varepsilon \hbar \omega) \sim \alpha$. When $q \sim k_F$, the hydrodynamic approach is not adequate to describe the plasmon dispersion relation quantitatively. The key ingredient of a fully quantum-mechanical calculation of plasmon modes is the so-called retarded (or causal) density-density response function[40] $\chi_{nn}(q,\omega)$, which, in the framework of Random Phase Approximation (RPA) and linear response theory, is given by $\chi_{nn}^{RPA}(q,\omega) = \chi_{nn}^{(0)}(q,\omega) / \varepsilon(q,\omega)$, where $\chi_{nn}^{(0)}(q,\omega)$ is the well-known[44,45,48,49] response function and $\varepsilon(q,\omega)$ is the RPA dynamical dielectric function. The function $\chi_{nn}^{(0)}(q,\omega)$ is usually referred to as the "Lindhard function". Note that RPA is not exact as shown by recent experiments[50]. Corrections to the dispersion relation eq. (5) are given in ref. [51]. So far, we have discussed plasmons in doped graphene sheets at zero temperature and in the absence of disorder. Finite-temperature and disorder effects can be incorporated rather easily in the RPA theory by employing the appropriate Lindhard function at finite temperature[52] and in the relaxation time approximation[53], respectively. To the best of our knowledge, the intrinsic plasmon lifetime in a doped graphene sheet is yet to be calculated.

**The impact of screening due to a metal gate on Dirac plasmons**
Let us briefly discuss the impact of screening by a metal gate on the Dirac-plasmon dispersion eq. (5) in a doped graphene sheet. Neglecting effects of hybridization between graphene sheet and metal, we can describe a metal as a grounded conductor screening the Coulomb interactions between electrons in graphene which leads to a replacement $u_q \to U_d(q) = 2\pi e^2 (1 - \exp(-2qd))/q$, where $d$ is the distance between graphene and the gate. Since $U_d(q)$ is regular at $q=0$, we expect a gapless acoustic plasmon $\Omega_{ac}(q \to 0) = c_s q$ rather than an "unscreened plasmon" (5) with $\omega_{pl}(q) \propto \sqrt{q}$. The exact expression for $c_s$ is derived in ref.[54].

**Experimental observations of intrinsic plasmons in graphene**

Plasmons in 2DESs can be accessed by a variety of direct and indirect methods, including optical measurements, electron energy-loss spectroscopy (EELS), inelastic light scattering, angle-resolved photoemission spectroscopy (ARPES) and scanning tunnelling spectroscopy. Several EELS experiments have been performed on exfoliated graphene sheets[55] and on epitaxial graphene samples[56], which showed that Dirac plasmons in graphene on SiC are strongly hybridized with the surface-optical phonons of the SiC substrate[57]. Dirac plasmons have also been probed by directly engineering their coupling to infrared light in a number of intriguing ways[58-63].

Ju *et al.*[58] have studied large-area CVD grown arrays of graphene microribbons demonstrating that infrared light polarized perpendicular to the ribbon axis is able to excite plasmon resonances of the confined MDF gas (see Fig. 3b). In these experiments, an ion gel was used for gating with the induced carrier concentration at the level of $10^{13}$ cm$^{-2}$ which allowed one to study the terahertz region. The authors of ref.[57] have shown that plasmon excitations in the graphene micro-ribbon array can be varied by electrical gating, Fig. 3c. They also demonstrated that the plasmon frequency scales like $W^{1/2}$, where $W$ is the width of the ribbon, and like $n^{1/4}$, in agreement with the bulk RPA prediction in eq. (5). The graphene plasmonic 6mm-long waveguides based on microribbons allowed a transmission of 2.5 Gbps optical signals with the averaged extinction ratio of 19 dB at a wavelength of 1.31 μm[63].

The direct interaction of localized graphene plasmons with infrared light has been demonstrated in ref.[60] (Fig. 3d). The authors used the stacked micro-disc geometry which allowed plasmon tuneability by changing the disk diameter, the number of stacks, the filling number and the gating (Fig. 3d). It was found that the collective oscillation of Dirac fermions is unambiguously quantum-mechanical with $n^{1/4}$ scaling of plasmon frequency. In addition, stacked graphene micro-discs were shown to realise an electromagnetic radiation shield with 97.5% effectiveness, a tuneable far-infrared notch filter with 8.2 dB rejection ratio, and a tuneable terahertz linear polarizer with 9.5 dB extinction ratio[60].

Perhaps the most striking were the recent observations of intrinsic graphene plasmons by Fei *et al.*[61] and Chen *et al.*[62] where the authors used the tip of an AFM probe in a scattering-type scanning near-field optical microscopy (s-SNOM) setup to launch and image Dirac plasmons in real space. A schematic picture of the experiments carried out and some of the findings are summarized in Fig. 4a. Infrared nano-imaging revealed that the plasmon wavelength compression ratio $\lambda_0 / \lambda_{pl}$ can reach 40 and that the plasmon in confined geometries can be tuned by gating. The strong confinement of graphene plasmons allowed one to use a single molecule defect in graphene as an atomic antenna in the petaHertz frequency range[64].

Crassee *et al.*[65] found that nanoscale morphological defects, such as atomic steps and wrinkles, in epitaxial graphene on SiC are responsible for a strong THz plasmonic peak in the optical response. Plasmons in epitaxial graphene can thus couple to THz light in the absence of artificial lithographic patterning.

The possibility to change the plasmon frequency easily by employing gates is important. However, Dirac plasmons are also fascinating for other reasons: i) the Dirac-plasmon dispersion suffers renormalizations due to exchange and correlation effects even in the long-wavelength limit; ii) Dirac plasmons can be manipulated by employing surface-science techniques; iii) it is easy to couple Dirac plasmons to light provided that the momentum mismatch between the latter and the former is compensated by a sufficiently "sharp" object (a periodic pattern, the tip of an AFM probe, a nanoscale defect, etc); iv) the lifetime of Dirac plasmons is predicted to be much longer than that of plasmons in metals or semiconductors.

**Plasmarons**

Condensed matter physics is rich in examples of "composite" quasiparticles, which are often bound states of bare electrons or holes with collective excitations of a many-particle medium, e.g., surface plasmon-polaritons. In graphene, composite quasiparticles formed by the coupling of elementary charges with plasmons are called plasmarons. Their existence has been predicted[46,66] and then demonstrated[67,68] in doped graphene sheets. A plasmaron can be regarded as a "dressed" electron (or a hole) interacting with plasmonic vibrations as suggested by Lundqvist[69]. Plasmarons can be observed by optical and tunnelling spectroscopies. Recently, Bostwick *et al.*[67] and Walter *et al.*[68] conducted a series of high-resolution ARPES experiments on high quality graphene sheets grown on SiC demonstrating a massive reconstruction of the MDF chiral spectrum near the Dirac point of doped samples. A representative set of ARPES data for the spectral function of the electron gas in quasi-freestanding graphene sheet is shown in Fig. 4b.

**Plasmon-electron interactions in graphene**

The particle-hole continuum of a non-interacting gas of MDFs is represented by the imaginary part of the Lindhard function which (for an undoped MDF system) is given by

$$\text{Im}[\chi_{nn}^{(0)}(q,\omega)]|_{undoped} = -\frac{q^2}{4\hbar} \frac{\Theta(\omega - v_F q)}{\sqrt{\omega^2 - v_F^2 q^2}}. \tag{6}$$

The main features here is the divergence of $\left|\omega^2 - v_F^2 q^2\right|^{-1/2}$ that occurs near the "light cone" $\omega = v_F q$ and the relatively weak weight at the lower limit of the $q < k_F$ inter-band particle-hole continuum. The divergence arises from the linear MDF dispersion, which places the maximum intra-band particle-hole excitation energy at $v_F q$ for all $k$. The plasmon excitation of the Dirac sea remains remarkably well defined even when it enters the inter-band particle-hole continuum. Interactions between quasiparticles and plasmons are stronger in the 2D MDF system than in an ordinary non-relativistic 2DES.

In a doped graphene sheet we need to take into account the dynamically screened RPA interaction $W(q,\omega) = u_q / \varepsilon(q,\omega)$. Calculating the "self-energy" of the quasiparticles we find that at some specific wave number, the bare quasiparticles velocity equals the plasmon group velocity and the charge carrier scatters into a resonance consisting of a quasiparticle "traveling together" with (i.e. strongly coupled to) an undamped plasmon excitation, a plasmaron. This happens at $\omega(q) \approx v_F^*(q - k_F)$, with $v_F^* > v_F$ being the enhanced quasiparticles velocity[70]. Plasmaron features can also be observed by tunnelling spectroscopy[71,72]. More details on the properties of electron-electron interactions and its influence on graphene optics can be found in the review[73].

## Graphene based plasmonics - hybrid devices

Graphene is a versatile, broadband, adjustable and tuneable optical material. However, direct applications of graphene in optics and photonics suffer from graphene's relatively inefficient interaction with light. While one might argue that 2.3% of absorption of light by a single atomic layer is actually a large number, in order to achieve effective optical modulators and photocells it is necessary to enhance light-matter interactions in a graphene sheet. The combination of graphene with conventional plasmonics based on noble metals[74,75] could, therefore, be beneficial for both field of investigations: i) the plasmonic nanostructures could enhance the optical properties of graphene (stronger Raman signature, more effective graphene plasmonic photocells) and ii) the graphene could be applied to effectively influence the optical response of plasmonic nanoarrays (for optical modulators and sensing) leading to graphene based active plasmonics.

### Raman scattering in graphene enhanced by near-fields of plasmonic nanostructures

Metallic nanostructures excited by light often demonstrate localised surface plasmon resonances (LSPR) which are characterized by strongly enhanced near-fields produced by charges stopping at the surface of metal. Since light interactions with graphene are determined by the local electromagnetic fields (induced on a graphene sheet), one can effectively increase the light-matter interactions with graphene by placing metal nanostructures close to graphene. This strategy was employed in[76-79], where the LSPR of metallic nanodots were used to significantly increase the Raman intensity, as demonstrated in Fig. 5a. Graphene provides the ideal prototype test material to investigate surface enhanced Raman spectroscopy (SERS)[80]. Its Raman spectrum is well-known[81], graphene samples are entirely reproducible and can be made virtually defect free. It was shown[76] that the 2D nature of graphene allows a closed form description of the Raman enhancement based on the electromagnetic mechanism, in agreement with experiments. The development of a generic procedure for graphene transfer on top of a prefabricated plasmonic nanostructure[77] offers the possibility that graphene could become a test object of choice for studying and quantifying field amplification in the conventional plasmonics.

### Plasmonic enhancement of photovoltage in graphene

Near-field enhancement by plasmonic nanostructures was used to significantly improve the efficiency of photo-voltage conversion in graphene[82] and to achieve a spectral selectivity that enables multicolour photodetection[83], see Fig. 5b. Graphene-based photodetectors have excellent characteristics in terms of quantum efficiency and reaction time[84], because of the very large room-temperature mobility and high Fermi velocity of its charge carriers. A combination of graphene with plasmonic nanostructures allows one to enhance the photovoltage by 15-20 times without compromising the operational speed of a graphene photodetector. It worth noting that the exact mechanism for light to current conversion is still debated[85,86]. Recently, the importance of thermoelectric effects have been demonstrated in direct measurements[87] of the symmetry of six-fold photovoltage patterns as a function of bottom- and top-gate voltages (Fig. 5c). These patterns, together with the measured spatial and density dependence of the photoresponse, provide evidence that at low temperatures nonlocal hot carrier transport dominates the intrinsic photoresponse in plain graphene with non-structured electrodes[87].

The important element of a graphene based photodetector is a p–n junction usually required to separate the photo-generated electron–hole pairs. Such p–n junctions are often created close to the contacts, because of the difference in the work functions of metal and graphene[88]. The geometrical position (and "strength") of the p-n junction can be varied by gating which can be employed to modulate the photoresponse of graphene (Fig. 5d). The presence of electric contact between graphene and plasmonic nanostructures is important as it could result in resistive coupling of LPRS[89] and lead to additional doping of the graphene sheet. In this respect, the vertical geometries, shown in Fig. 2c, are more beneficial for light harvesting[90].

**Graphene based plasmonics for modulation and sensing**
Graphene tuneability by gating is vital for the field of active plasmonics. Active optical elements are of great importance in different areas of science and technology starting from ubiquitous displays and finishing with high tech frequency modulators. Despite a great deal of progress in optical disciplines, active optics still relies heavily on either liquid crystals, which guarantee deep modulation in inexpensive and small cells but are quite slow, or non-linear optical crystals, which are fast but quite bulky and expensive. For this reason, the development of inexpensive, fast and small active optical elements would be of considerable interest.

Recently, plasmonic metamaterials have established themselves as a versatile tool for creating new optical devices. Their optical properties can be easily controlled by changing the electric coupling between plasmonic nanoresonators which constitute a nanomaterial. To achieve such control one can use the extraordinary optical, electrical and mechanical properties of graphene. The combination of graphene with plasmonics could results in fast, relatively cheap and small active optical elements and nano-devices. There are two main challenges on this route: i) to combine graphene with plasmonic elements and ii) to achieve an effective control of graphene properties and optical response of hybrid optical devices.

Papasimakis *et al.*[91] applied a low pressure CVD process to grow graphene[92] on polycrystalline Cu foils and then utilized a PMMA wet transfer procedure to transfer graphene on the top of prefabricated plasmonic metamaterials with sizes of 20 $\mu m^2$ (Fig. 6a). The authors found that the graphene changed the spectral position of the plasmonic resonances of the metamaterial as well as the absolute value of spectral transmission, see Fig. 6b. Simple FDTD calculations[93] provided general support for the experimental data, although the details of graphene resistive coupling to plasmons and graphene doping by the metal are still lacking.

Mechanically exfoliated graphene and a wet PMMA transfer procedure were used to place graphene on top of regular plasmonic nanoarrays[77,94] with sizes of 200 $\mu m^2$. These plasmonic nanoarrays possess ultra-narrow diffractive coupled plasmon resonances[95] (or geometric resonances) which show extremely high phase sensitivity to the external environment[96]. The presence of graphene strongly influenced both the position of coupled resonances and other optical properties of the samples. To modify its properties, reversible graphene hydrogenation was performed in which an areal mass sensitivity limit of 0.1 fg/mm$^2$ was recorded[94]. This is 4 orders of magnitude better than the areal sensitivity of the surface plasmon resonance technique[97] and thus opens up the possibility of realising single-molecular label-free detection. Due to the well-defined 2D geometry and possibility to monitor the hydrogenation level using

Raman[98], graphene could become a test object of choice for measuring areal mass sensitivity in plasmonic nano-sensors.

The ultimate goal for hybrid graphene/plasmonic elements is to achieve light modulation using graphene gating, see Fig. 6d where a hypothetical plasmonic interferometer[99] governed by graphene is shown. The presence of metallic nanostructures makes the problem of gating quite complicated due to electrostatics and electric discharge. The problem of graphene gating in hybrid plasmonic nano-devices is currently under intense investigation.

## Perspective

Graphene, its derivatives and other single atomic layer materials could become building blocks of new generations of optical devices. A new multilayer LEGO game in flatland now contains all kinds of optical materials: dielectric, semiconducting, semi-metallic, metallic, gapless, with small and large optical gaps, etc. One may hope that this LEGO game will bring about new optical applications (efficient photocells, ultrafast optical modulators, graphene-based 2D lasers) in the near future. Graphene plasmons are also coming of age and show very attractive features which include extremely high localization, strong confinement, efficient and strong light-matter interactions, relatively large life-times, tuneability and electric controllability. Light manipulation with intrinsic graphene plasmons and the accessibility of quantum optical regimes promise a revolution in optoelectronics and optical computing. A combination of graphene with conventional plasmonic nanostructures and metamaterials will provide ultrasensitive chemical and biosensors, new non-linear optical elements and effective photodetectors. We envisage that graphene could become a test object of choice for quantifying field amplification in plasmonics and measuring areal sensitivity of plasmonic devices. At the same time, the rich and diverse properties of graphene quasiparticles and their interaction with plasmonic vibrations promise new discoveries in fundamental physics.

**Acknowledgements:** We thank A. Geim, V. I. Fal'ko, M. I. Katsnelson, R. Asgari, R. Fazio, F. Guinea, A. H. MacDonald, V. Pellegrini, E. Rotenberg, F. Taddei, G. Vignale for very useful and stimulating conversations. M.P. was supported by the Italian Ministry of Education, University, and Research (MIUR).

**Figure captions.**

Figure 1. Building blocks of flatland optics. **a**, Left: the optical conductivity of pristine graphene (blue and cyan lines – with and without electron-electron interactions), pristine bilayer graphene (red line), doped graphene (pink line), doped bi-layer graphene (dark green line) and fluorographene (black line). Inset shows zoom-in at the low energy range of spectrum. Right: absorption spectra of a single and double layer of molybdenum disulphide (left axis, normalized by the number of layers, black lines) and the corresponding PL spectra (right axis, normalized by the intensity of the peak A, red lines). The spectra are displaced along the vertical axis for clarity. The green line shows the spectral position of the excitation wavelength. **b**, Pauli blocking of photon absorption in graphene. **c**, A sketch of graphene-based waveguide-integrated optical modulator, adapted from ref. [17]. Data for **a** are partially taken from ref. [31].

Figure 2. Multilayered flatland optics. **a**, Energy spectrum for pristine and doped bilayer graphene. **b**, Optical modulator based on tuneable bandgap in bilayer graphene. **c**, The hypothetical multilayer structure with the vertical charge separation. **d**, Cavity based graphene photodetector and graphene based integrated interferometer. Images reproduced with permission from: **b**, ref. [26]; **d**, adapted from ref. [39].

Figure 3. Intrinsic graphene plasmons. **a**, The schematics of plasmon excitation. **b**, Plasmon resonance in gated graphene micro-ribbon arrays. Top, from left to right: a top and a side views of a typical graphene micro-ribbon array, AFM image of a graphene micro-ribbon array sample, gate-dependent electrical resistance of this graphene micro-ribbon array. Bottom: gate-induced change of transmission spectra. **c**, Control of plasmon resonance through electrical gating and micro-ribbon width. Top: spectra as a function of gate voltage. Bottom: AFM images of samples, left, transmission spectra for the corresponding samples for the same doping concentration, right. **d**, Transparent graphene plasmonic devices. Top: SEM image (false colour) of a stacked graphene/insulator microdisc array. Bottom: Extinction in transmission in stacked plasmonic devices with one, two and five graphene layers. Figures reproduced with permission from: **b** and **c**, ref. [58]; **d**, ref. [60].

Figure 4. Launching and imaging graphene plasmons. **a**, Top left: schematic of an infrared nano-imaging experiment. The blue and green arrows label the incoming and back-scattered light, respectively. Top right: images of various interference patterns close to graphene edges (blue dashed lines), defects (green dashed lines), or boundaries between single-layer and bilayer graphene (white dashed line). Bottom: controlling the plasmon wavelength in scattering experiments. **b**, Plasmaron satellite bands in the ARPES spectrum of a graphene on SiC. (A) The Dirac energy spectrum of a graphene in a non-interacting picture. (B and C) Experimental spectrum of doped graphene perpendicular and parallel to the ΓK direction. The dashed lines are guides to the dispersion of the observed hole and plasmaron bands. The red lines are at $k$ = 0. (D to G) Constant energy cuts of the spectral function at different binding energies. (H) Schematic Dirac spectrum in the presence of electron-electron interactions. The scale bar in (C) defines the momentum length scale in (B) to (G). Figures reproduced with permission from: **a**, refs. [61], [62]; **b**, ref. [67].

Figure 5. Hybrid graphene plasmonic devices. **a**, Top left: SEM images of SERS sample. Top right: the total (patterned) enhancement factors for the G and 2D peaks. The dotted line is the corresponding interference (unpatterned) enhancement factor. Bottom left: Schematics of plasmonic metamaterial and graphene. Bottom right: Raman spectra of graphene placed on top of gold nanodots measured at 514 and 633 nm. **b**, Top left: SEM images of the graphene photodetectors with plasmonic nanostructures. Top right: the corresponding Resistance and photovoltaic characteristics. Bottom: multi-colour photodetection using graphene devices coupled with different plasmonic nanostructures with SEM images and the corresponding photoresponse. **c**, Device geometry, band structure, and optoelectronic characteristics of the graphene p-n junction. (A) Optical microscope image of the dual-gated device. (B) Resistance versus VBG and VTG at VSD = 1.4 mV and T = 175 K. (C) Experimental schematic (top) and schematic of monolayer graphene band structure in the p-n junction (bottom). (D) Spatially resolved photocurrent map at T = 40 K with laser wavelength 850 nm and optical power = 50 mW. (E) Photocurrent line trace taken at dashed line in (D). **d**, The schematics of charge separation for graphene doped by metallic contacts. Figures reproduced with permission from: **a**, refs. [76], [77]; **b**, refs. [82], [83]; **c**, ref. [87].

Figure 6. Hybrid graphene plasmonic devices. **a**, Top: schematic of the fabricated complementary split-ring metamaterial. Middle: a microscope image after deposition of graphene. Bottom: Raman spectrum corresponding to the marked region. **b**, Top: experimental transmission spectrum of an array before (dashed black line) and after (red solid line) deposition of graphene for unit cell size D = 711nm as shown in the inset. Middle: wavelength shift of plasmon resonances due to graphene as a function of unit cell size. Bottom: electric field maps at the plasmon resonances. **c**, Evaluation of sensitivity for singular-phase plasmonic detectors. Top left: ellipsometric spectra in the region of the collective plasmon resonance for the pristine double-dot array (black curve) and with graphene transferred on top (red). The inset shows the entire spectrum for the pristine case. Top right: Evolution of reflection during hydrogenation and annealing: the red curve corresponds to initial spectra; green - 20min of hydrogenation, blue - 60min, black - after annealing. Bottom left: the ratio of the amplitudes of D and G peaks in graphene as a function of hydrogenation. Bottom right: $\Psi$ and $\Delta$ for the case of 1% hydrogenated and pristine graphene as a function of wavelength (incident angle of 70). **d**, A hypothetical graphene based active plasmonic interferometer. Figures reproduced with permission from: **a**, **b**, ref. [91]; **c**, ref. [94].

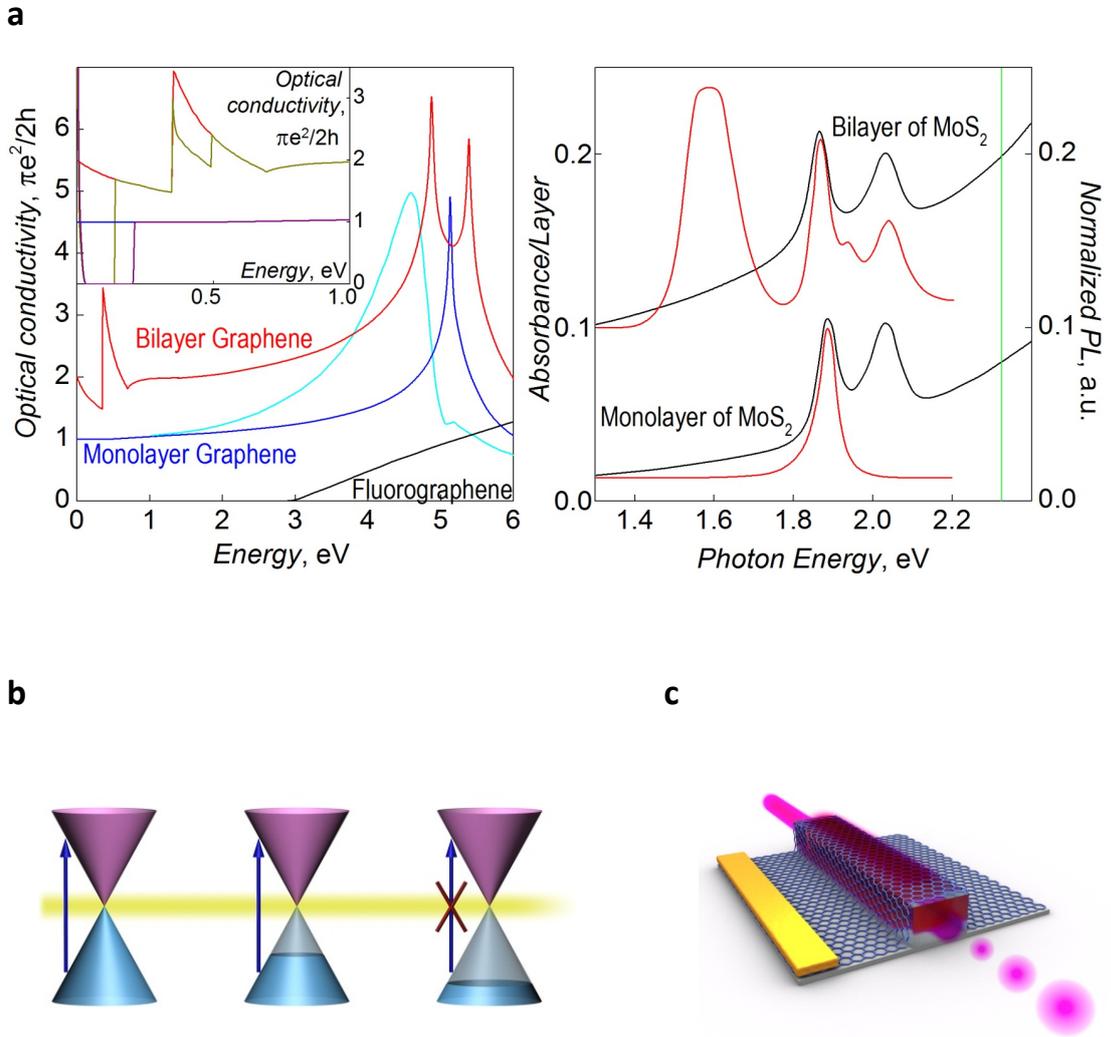

Figure 1

a 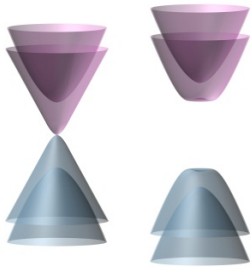 b 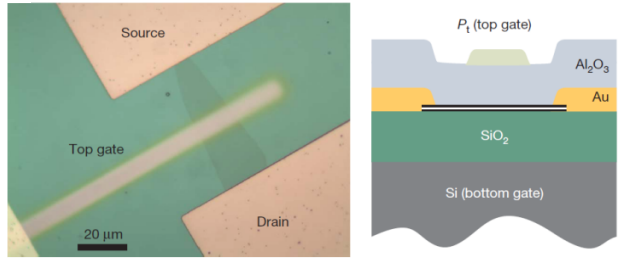

c 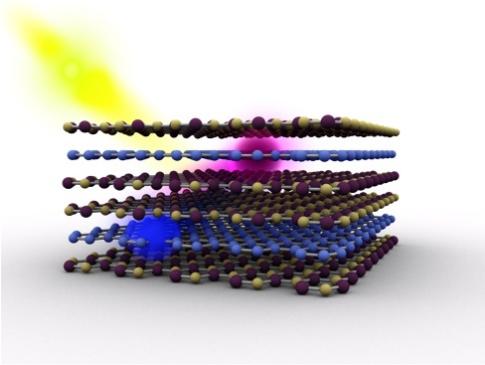 d 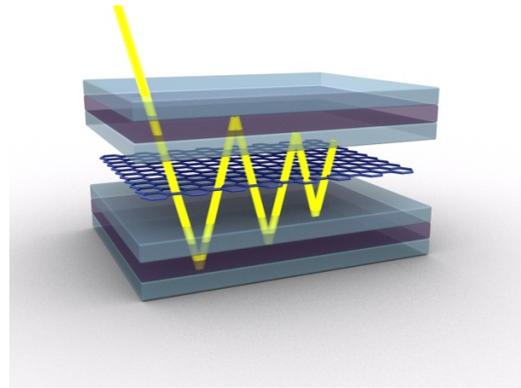

Figure 2

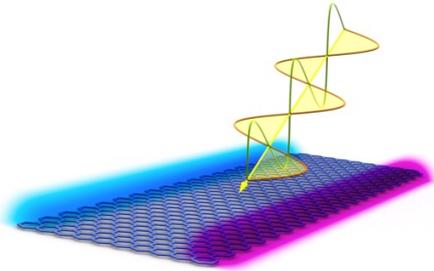
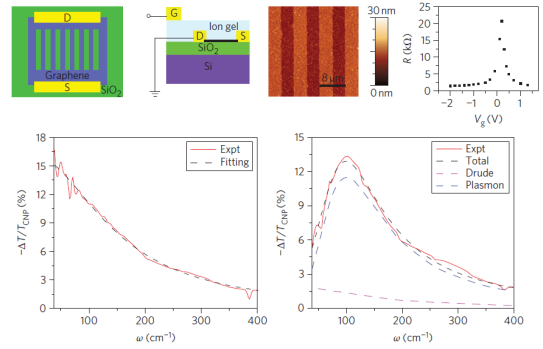
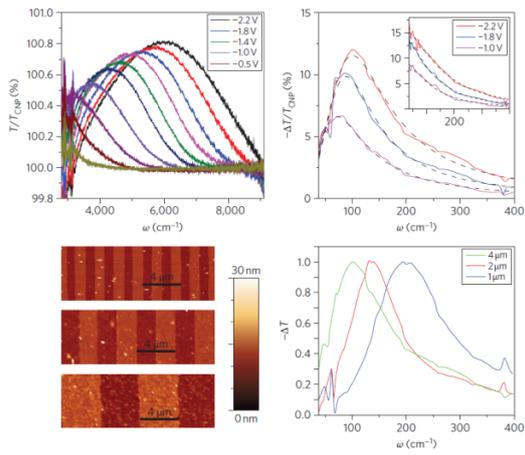
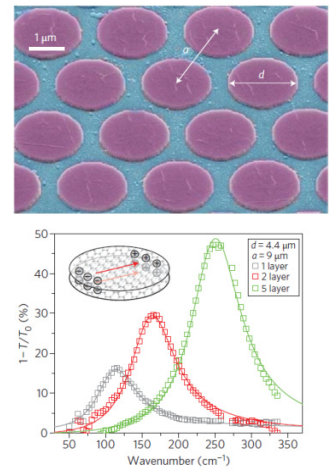

Figure 3

**a**

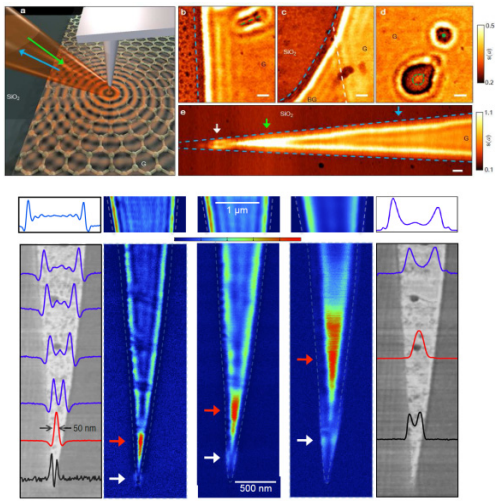

**b**

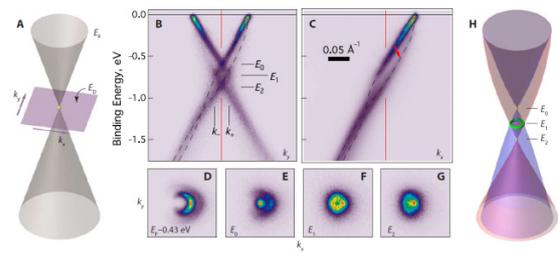

Figure 4

a 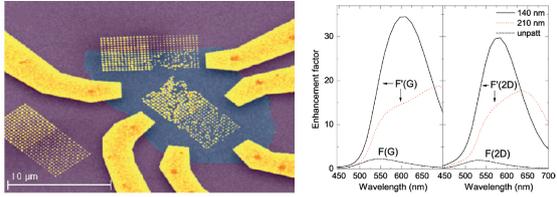 b 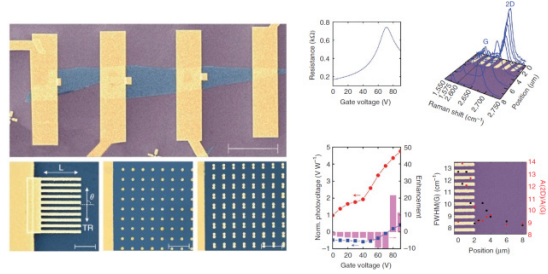

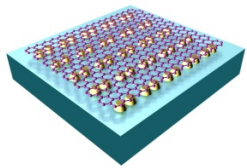 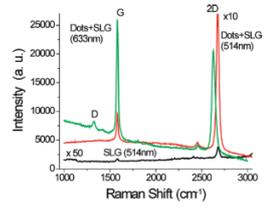 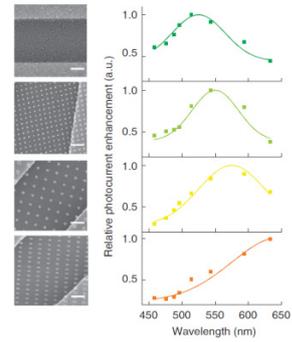

c 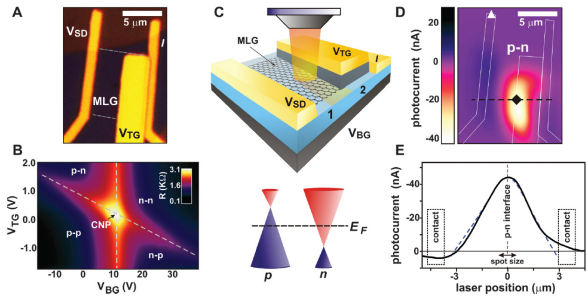 d 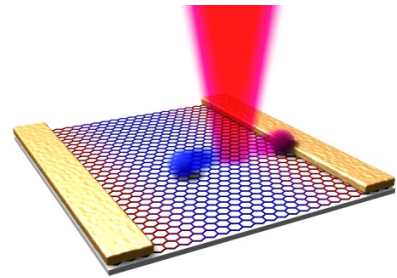

Figure 5

a

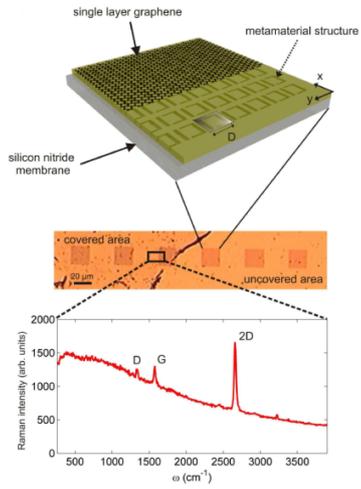

b

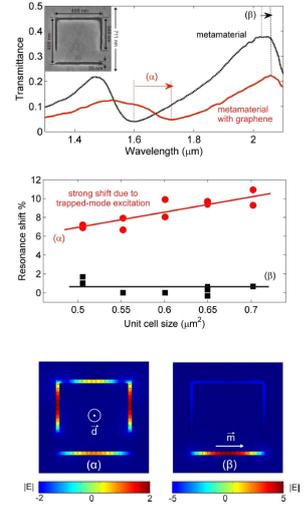

c

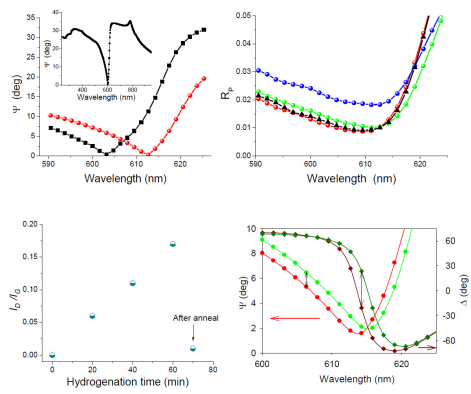

d

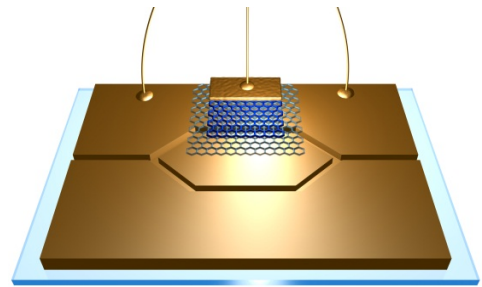

Figure 6